\documentstyle[tighten,prl,epsf,floats,aps]{revtex} 
\date{June 24, 1997, submitted to Phys. Rev. Lett.} 
\begin{document}

\twocolumn[
\hsize\textwidth\columnwidth\hsize\csname @twocolumnfalse\endcsname

\draft
\preprint{\today}
\title{New Far Infrared Vibrational Mode  in Zn Doped CuGeO$_3$}

\author{J. J. McGuire$^{1}$, T. R\~o\~om$^{1,}$\cite{Toomas},  T. E. Mason$^{2}$, 
T.~Timusk$^{1}$} 
\address{$^{1}$ Department of Physics and 
Astronomy, McMaster University, Hamilton, Ontario L8S 4M1, Canada} 
\address{$^{2}$ Department of Physics, University of Toronto, 
Ontario M5S 1A7, Canada} 

\author{H. Dabkowska} 

\address{Brockhouse Institute for Materials Research, McMaster 
University, Hamilton, Ontario L8S 4M1, Canada} 

\author{S. M. Coad, D. McK. Paul} 
\address{Department of Physics, University of Warwick, Coventry CV4 7AL, UK}

\maketitle 

\begin{abstract}
We report on far infrared measurements on Zn and Si doped crystals of 
the spin-Peierls compound CuGeO$_3$.  Zn doping has the effect of 
introducing several new absorption lines, polarized in the $ab$-plane, 
between 5 and 55~cm$^{-1}$.  The intensity of the absorption grows 
with Zn concentration but saturates above 2\,\% Zn.  One line at 
10~cm$^{-1}$ loses intensity above 4~K, and a second line at 20~cm$^{-1}$ is 
absent at low temperatures but grows to peak at about 40~K in 
agreement with a three level model with two excited states 10 and 
30~cm$^{-1}$ above the ground state.  As the doping is increased these lines 
broaden, and a temperature independent absorption develops over the entire 
range from 5 to 55~cm$^{-1}$.  These features are magnetic field independent up 
to 16~T and are absent in Si doped samples.  We suggest the new 
absorption is due to localized lattice modes of the zinc ion and the 
surrounding GeO$_4$ tetrahedra. 
\end{abstract} 

\pacs{PACS numbers: 75.50.Ee, 78.30.Hv, 63.20.Pw} 
]


The quasi-one-dimensional antiferromagnet CuGeO$_3$ is the first 
inorganic compound to show a spin-Peierls (SP) 
transition\cite{Hase93}. At $T=14$~K a gap opens in the excitation 
spectrum of the Cu $S=\frac{1}{2}$ spins as a result of magneto-elastic 
coupling. In general, such transitions involve a lowering of magnetic 
energy at the expense of lattice energy. Structures 
where large tightly bound blocks can move as units, such as the 
organic charge transfer complexes, exhibit the spin-Peierls 
phenomenon. Similar structural units occur in oxides in the form 
of rigid GeO$_4$ or SiO$_4$ tetrahedra bonded weakly to each other to 
form easily distorted structures\cite{Hammond97}. It has been found 
that in CuGeO$_3$ the SP transition involves both a dimerization of the 
copper ions along the {\bf c}-axis and a rotation of the GeO$_4$ 
tetrahedra about the {\bf c}-axis\cite{Hirota94}. In this paper we 
present the results of a far infrared transmission study of Zn and Si 
doped CuGeO$_3$. Our results show that Zn substitution for Cu induces new 
absorption lines at 10 and 20~cm$^{-1}$.   We show that these transitions 
result from a very unusual, highly anharmonic localized vibrational mode. 

\begin{figure}[h]
\leavevmode
\epsfxsize=\columnwidth
\centerline{\epsffile{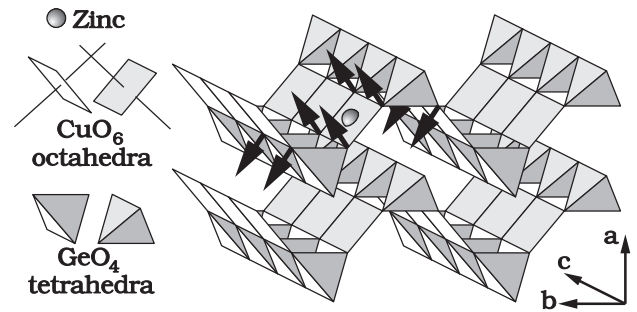}}
\vspace{0.1in}
\caption{Lattice structure of CuGeO$_3$. Arrows show possible motion of the rigid GeO$_4$
tetrahedra ascociated with a Zn-induced local vibrational mode.
}
\label{figure1}
\end{figure}

The lattice structure of CuGeO$_3$ is shown in Fig.~1.
It consists of chains of relatively rigid GeO$_4$ tetrahedra along 
the {\bf c}-axis. Cu$^{2+}$ ions are positioned between these chains and 
form chains of CuO$_6$ octahedra which share the GeO$_4$ oxygens. 
Adjacent octahedra share two oxygens which provide a superexchange 
path for the antiferromagnetic interaction between Cu$^{2+}$ ions.

The far infrared absorption spectrum of pure CuGeO$_3$ has a weak 
absorption line in the SP phase at 44~cm$^{-1}$ \cite{Loosdrecht96} which 
corresponds to the SP gap energy ($\approx5.5$~meV) at 
${\bf Q} =[0,0,0]$ of the magnetic Brillouin zone as determined by electron 
spin resonance\cite{Brill94}. The line splits in a magnetic field as 
expected for a transition from the ground singlet state to the 
excited triplet state. The other low-lying spectral feature is a 
weak phonon at 48~cm$^{-1}$ suggested to correspond to a displacement of 
the four planar oxygens in the direction of the apical 
oxygens in the CuO$_6$ octahedra\cite{Popovic95}.

Doping of CuGeO$_3$ with Zn or Si is particularly interesting because
it results in an antiferromagnetic (AF) phase that 
seems to coexist with the SP phase in a range of temperatures and 
doping levels in Cu$_{1-x}$Zn$_{x}$GeO$_3$\cite{Lussier95,Sasago96,Hase96a,Fronzes97,Martin97} 
and in CuGe$_{1-x}$Si$_{x}$O$_3$\cite{Renard95,Regnault95,Poirier95}. 
There are, to our knowledge, no measurements of the far infrared spectrum 
of the doped systems of CuGeO$_3$.


We studied several single crystals including an undoped sample, a Si 
doped sample with $x=0.003$, and a series of Zn doped samples with 
$x=0.003, 0.01, 0.02$ and $0.05$. The N\'eel and SP transition 
temperatures, $T_N$ and $T_{SP}$, were determined from magnetic susceptibility 
measurements on the same crystals.  The $x=0.01$, 0.02 and 
undoped samples were grown by a floating zone technique and the 
$x=0.003$ and 0.05 samples by a self flux method.  We used either 
atomic emission spectroscopy or mass spectroscopy to determine the Zn 
and Si concentrations and found good agreement with phase 
diagrams\cite{Sasago96,Martin97,Renard95} for $T_{SP}$ and $T_N$ {\it 
vs.} concentration $x$.  

The far infrared measurements were done with a home-built polarizing 
Fourier spectrometer\cite{spec} and a new magnet insert with an 
{\it in situ} $^3$He cooled ($T=0.3$~K) silicon bolometer.  
Polarization-dependent transmission measurements were done in 
magnetic fields up to 16~T from 3 to 100~cm$^{-1}$ and in the temperature 
range from 1.2 to 100~K.  Sample sizes were about $0.6\times 2.5 
\times 4$~mm$^3$ in the directions of the {\bf a}, {\bf b} and {\bf c} 
crystallographic axes\cite{Vollenkle67} respectively. The {\bf k}-vector 
of the light was aligned along the {\bf a}-axis ({\bf k}$\parallel${\bf a}). 
To establish the symmetry of the transitions, additional measurements were done with 
{\bf k}$\parallel${\bf b} for one of the Zn doped samples ($x=0.01$). The dc 
magnetic field was in the direction of light propagation (Faraday 
geometry), and most of the spectra were measured at 1~cm$^{-1}$ resolution. 
To calculate the absorption coefficient $\alpha$ from the 
transmittance it is necessary to estimate the refractive index. We 
used the interference fringes from a thin sample with parallel faces 
to do this.  A constant value of $n=3$ was determined for the entire 
range from 3 to 100~cm$^{-1}$. 


The absorption spectra for several samples at 1.2~K are shown in 
Fig.~2.  The sample doped with $x=0.003$ Si is very similar to the 
undoped sample (not shown) in that the only far infrared features are 
the line at 48~cm$^{-1}$ which has been identified as 
a $B_{3u}$ phonon\cite{Popovic95}, and the shoulder at 
44~cm$^{-1}$ due to the singlet to triplet transition\cite{Loosdrecht96}.  
In contrast, the $x=0.003$ Zn sample shows, in 
addition to these features, a line at 10~cm$^{-1}$ which is not present in 
the Si doped sample (or in the undoped samples).  Increased doping 
results in a broadening of the 10~cm$^{-1}$ line and the development of a 
broad absorption band between 5 and 55~cm$^{-1}$.  The variation of the 
spectral weight\cite{units} from 3 to 60~cm$^{-1}$ (background and phonon 
line subtracted) is shown in the inset to Fig.~2 as a function of 
concentration.  The dependence on Zn concentration 
appears to be linear at low concentrations but quickly saturates 
above $x=0.02$.

Absorption spectra for the $x=0.003$ Zn doped sample were also obtained 
at higher temperatures.  While the shoulder at 44~cm$^{-1}$ disappears 
above the SP transition temperature as expected, the 10~cm$^{-1}$ line and 
another line at 20~cm$^{-1}$ show a temperature dependence that has no 
obvious relationship to the SP transition.  The spectra are shown in 
the upper panel of Fig.~3, and the lower panel shows the spectral 
weight of the 10 and 20~cm$^{-1}$ lines as functions of temperature. 
The 10~cm$^{-1}$ line loses intensity as temperature is increased, but the 
20~cm$^{-1}$ line, which is absent at low temperatures, grows in intensity 
to peak around 40~K.  Both lines shift to higher frequencies with temperature 
(inset to Fig.~3), and the 20~cm$^{-1}$ line broadens significantly. 
Although the 10 and 20~cm$^{-1}$ lines are broader in the more highly doped
samples, they still show this temperature dependence. 
The broad absorption from 5 to 55~cm$^{-1}$ in samples with $x>0.003$ is 
temperature independent up to 100~K.

\begin{figure}[t]
\leavevmode
\epsfxsize=\columnwidth
\centerline{\epsffile{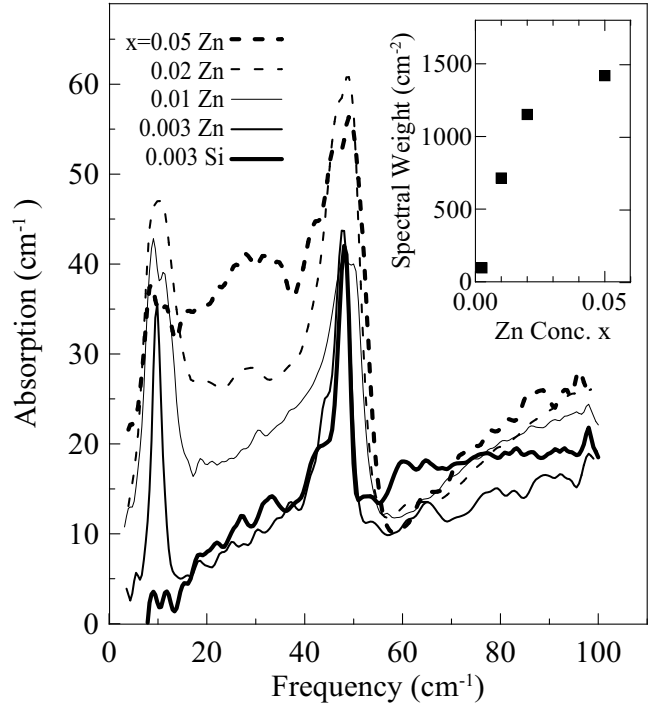}}
\vspace{0.15in}
 \caption{Doping dependence of far infrared absorption
of CuGeO$_3$ crystals at 2~K
with $b$-polarization and {\bf k}$\parallel${\bf a}.
The inset shows spectral weight from 3 to 60~cm$^{-1}$
as a function of Zn concentration.
The spectral weight does not include the background which
is taken to be the Si doped sample spectrum.}
 
\label{figure2}
\end{figure}

The temperature dependence of the 20~cm$^{-1}$ line suggests that it is a 
transition from an excited state.  A simple three level model with two 
excited states 10 and 30~cm$^{-1}$ above the ground state accounts 
for the temperature dependences of the line intensities 
in the lower panel of Fig.~3. The trends in the data are correctly 
reproduced, and the quality of the fit, though not 
particularly good, can be improved by adding higher energy levels to 
the model.  This points to a multilevel, very anharmonic 
excitation, which is made possible by the replacement of a Cu$^{2+}$ 
ion with Zn$^{2+}$, but not by the substitution of Si for Ge.

\begin{figure}[ht]
\leavevmode
\epsfxsize=0.85\columnwidth
\centerline{\epsffile{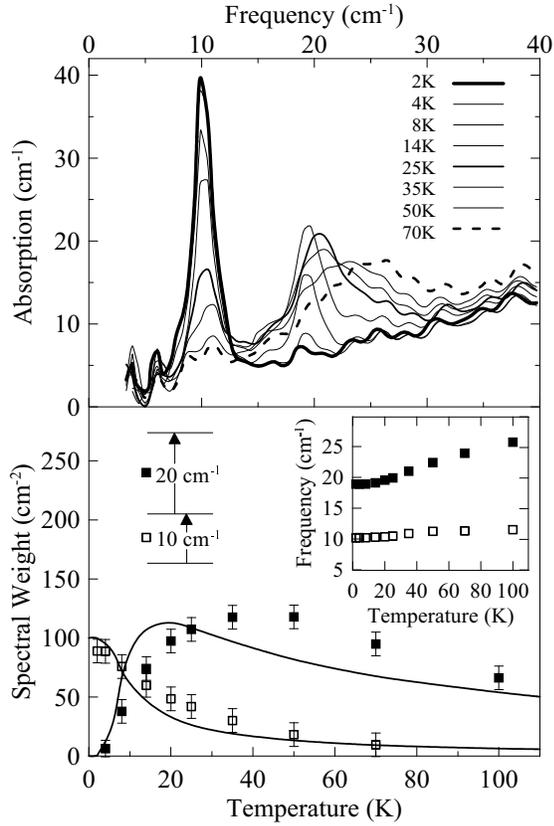}}
\vspace{0.1in}
 \caption{Temperature dependence of 10 and 20~cm$^{-1}$ lines
in the $x=0.003$ Zn doped sample.
The upper panel shows absorption at various temperatures
with $b$-polarization and {\bf k}$\parallel${\bf a}.
The lower panel shows the spectral weight of the lines
calculated using Lorentzian fits with frequencies shown in the inset.
Solid lines are fits to the simple three-level model:
$I_{10}(T)=A_{10} (1-e^{-\Delta_{10}\beta})/(1+e^{-\Delta_{10}\beta}+
e^{-(\Delta_{10}+\Delta_{20})\beta})$,
$I_{20}(T)=A_{20} e^{-\Delta_{10}\beta}(1-e^{-\Delta_{20}\beta})/(1+e^{-\Delta_{10}\beta}+
e^{-(\Delta_{10}+\Delta_{20})\beta})$,
where $\Delta_{10}$ and $\Delta_{20}$ are the temperature dependent transition frequencies
(inset), $A_{10}=100$ and $A_{20}=500$ are
adjustable parameters, and $\beta=(k_BT)^{-1}$.}

\label{figure3}
\end{figure}

The polarization dependence of the new features was also obtained.  
With {\bf k}$\parallel${\bf a}, the 10 and 20~cm$^{-1}$ 
lines as well as the 5 to 55~cm$^{-1}$ band were strongest when the 
electric field was along the {\bf b}-axis ($b$-polarization) and nearly 
absent with $c$-polarization.  When the $x=0.01$ Zn sample was 
oriented with {\bf k}$\parallel${\bf b}, the 
features were strongest with $a$-polarization.  Thus, if these are 
magnetic dipole transitions, they require that the magnetic field component 
of the light is along the {\bf c}-axis.  If they are electric dipole 
transitions, the electric field must be in the $ab$-plane and have 
components both in the {\bf a} and {\bf b} directions.

To exclude magnetic dipole transitions we compare the spectral weight 
of the Zn-induced absorption with that of a known magnetic dipole 
transition in our system.  An incommensurate phase exists in CuGeO$_3$ 
above $B_0\approx12$~T\cite{Hase93a,Kiryukhin96} where a magnetic 
dipole transition in the triplet ground state can be 
observed\cite{Loosdrecht96}. We find this transition (at 15~cm$^{-1}$ in 
Fig.~4) to be independent of polarization in the $bc$-plane 
($B_0=15$~T along the {\bf a}-axis) as expected for an $S=1$, $ \Delta 
M=\pm1$ transition and to have a spectral weight of 2.5~cm$^{-2}$.  
The spectral weight of the 10~cm$^{-1}$ line is 100~cm$^{-2}$, a factor of 40 
larger than the magnetic dipole transition. 
This discrepancy, which is even larger if we include the dilution due 
to low Zn concentration, in combination with the fact that the 
frequency of the Zn-induced absorption is not affected by magnetic 
field up to 16~T, leads us to conclude that it is not a magnetic 
dipole transition. 

\begin{figure}[t]
\leavevmode 
\epsfxsize=\columnwidth
\centerline{\epsffile{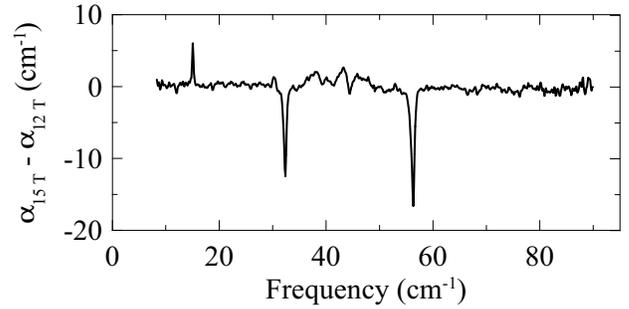}} 
\vspace{0.1in}
 \caption{Absorption in the SP phase at 12~T (negative features) subtracted from 
absorption in the incommensurate phase at 15~T (positive features) for the undoped sample
at 2.1~K and 0.2~cm$^{-1}$ resolution with $b$-polarization and {\bf k}$\parallel${\bf a}.}
\label{figure4}
\end{figure}

We can also rule out forbidden electronic electric dipole transitions 
such as the transition in the SP phase from the singlet ground state to the 
excited triplet state which produces the 44~cm$^{-1}$ line.  That line splits 
in a magnetic field as shown in Fig.~4 due to the splitting of the 
excited triplet state.  It is also ten times stronger than the magnetic 
dipole transition at 15~cm$^{-1}$, and as our study has shown, is induced when 
{\bf k}$\parallel${\bf a} with $b$-polarization only.  If 
the Zn induced features are electronic electric dipole transitions, 
it is unclear which states could be involved.  Since they are 
not split by a magnetic field, they are unlikely to involve 
transitions to or from triplet states.  In-gap states, predicted 
by Martins {\it et al.}\cite{Martins96}, are also excluded because 
they would be absent above the SP transition temperature. 

Local vibronic modes are another possibility.  For a pure Zn
translational mode for $x=0.003$ Zn 
we estimate a spectral weight of 530~cm$^{-2}$ from the 
plasma frequency $\omega_p=(4\pi n Z^2e^2/M)^{1/2}=13$~cm$^{-1}$ where $n$ 
is the Zn concentration, $M$ the mass of the Zn ion and $Z$ the 
effective charge which we take to be +2.  This is over 5 times larger 
than the observed spectral weight.  It is therefore likely
that the mode also involves motion of other atoms. 

The three level system shown in Fig.~3 can be modeled with a square 
well with a central barrier. Appropriate choice of parameters gives 
levels at 10, 30 and 55~cm$^{-1}$. Electric dipole transitions between 
levels of opposite parity will give lines at 10, 20, 25 and 55~cm$^{-1}$.  
The line at 25~cm$^{-1}$ is weak at the temperatures 
investigated because the population of the 30~cm$^{-1}$ level is small, and the 
55~cm$^{-1}$ line is weak because of  the 
small transition matrix element   from the ground state to the 
55~cm$^{-1}$ level. In general, the addition of the level at 55~cm$^{-1}$ will improve the 
agreement between the calculated and observed temperature dependance of the line
intensities shown in lower panel of Fig~3. 
The broadening of the 10 and 20~cm$^{-1}$ lines with increased 
doping can be explained as being due to long range of interaction between 
the defects.

The combination of a low intensity, a flat very anharmonic potential 
and a polarization in the $ab$-plane suggests that the lines at 10 
and 20~cm$^{-1}$ involve the Zn defect and the libration about 
the {\bf c}-axis 
of the surrounding four GeO$_4$ tetrahedra 
as shown in Fig.~1. In the undoped material the 48~cm$^{-1}$ phonon is 
a similar mode which is optically active\cite{Popovic95} and 
involves the motion of the Cu in the direction of the apical 
oxygens. With Cu replaced by Zn this mode would have a polarization with both 
{\bf a} and {\bf b} 
direction components and is consistent with the model of a potential 
well with a central barrier. Since Zn is known to favor a higher 
coordination number than Cu  it will have a tendency to move off 
center towards one of the apical oxygens.

Libration of GeO$_4$ tetrahedra is also a component of the {\it static} SP distortion in 
the undoped material\cite{Hirota94,Braden96}, but it is not optically active since 
it involves odd combinations of the rotations of the GeO$_4$ 
tetrahdra and no displacment of the copper in the $ab$-plane.
In our picture the Zn doping has the effect of moving the mode at 
48~cm$^{-1}$ to 10~cm$^{-1}$ by a dramatic reduction of the already weak 
restoring force for diplacements of the the planar oxygens in the  
direction of the apical oxygens around the Zn. 

In contrast to the large changes in dynamics by Zn doping, the 
replacement of Ge by Si will have the minor effect of
reducing the size of one of the tetrahedra  affecting four Cu 
sites equally and, by symmetry, giving rise to a low frequency optical
mode polarized in the {\bf b} direction. We have not observed such a mode 
in our Si doped samples from 3 to 100~cm$^{-1}$.

One feature which remains to be explained is the 5 to 55~cm$^{-1}$ band 
seen at higher doping levels.  
Its lower boundary is approximately defined by the 10~cm$^{-1}$ line,
and the upper boundary more or less coincides with the phonon at 
48~cm$^{-1}$.  The saturation of the absorption above a concentration of 
$x=0.02$ (inset to Fig.~2) is reminiscent of the phase 
diagram\cite{Sasago96,Martin97} in which $T_N$ and $T_{SP}$ 
stop changing at this same concentration.  This may be 
evidence for an upper limit for effective doping. Another 
possibility, suggested by the unusual flat shape of the band, is that 
it represents a disordered state of the local excitations as they begin 
to interact strongly at $x=0.02$ Zn and above.  This state 
may also be relevant to the saturation of $T_N$ and $T_{SP}$ 
at the same doping level.

In summary we have observed a low lying lattice mode in Zn doped 
CuGeO$_3$ which we idenify with the librational motion of GeO$_4$ 
tetrahedra combined with the displacment of Zn in a very flat 
potential with a central barrier. 


We acknowledge fruitful discussions with J. Barbier, D.B.~Brown,  
B.~Gaulin, and J.E.~Greedan. The work at McMaster University and 
 the University of Toronto was 
supported by  NSERC of Canada and The Canadian Institute for Advanced 
Research. Work on correlated magnetic systems at Warwick is supported 
by a grant from the EPSRC of UK.



\begin{thebibliography}{10}
\bibitem{spec}
Design of the far infrared spectrometer will be published later.
\bibitem{units}
We define spectral weight (in units of cm$^{-2}$) as $\int_0^\infty \alpha d\nu
  =(\pi^2/n)\omega_p^2$ where the absorption constant $\alpha$, the frequency
  $\nu$ and the plasma frequency $\omega_p$ are in units of cm$^{-1}$ and $n$
  is the index of refraction.
\end{thebibliography}

 \end{document}